\documentclass[aps,preprint,epsfig,rotate,showkeys]{revtex4}
\usepackage{graphicx}
\usepackage{bm}
\usepackage{epsfig}
\usepackage{pdflscape}

\begin{document}
\title{On the $\beta^{-}$-decay in the ${}^{8}$Li and ${}^{9}$Li atoms}

\author{Mar\'{\i}a Bel\'en Ruiz}
\email[E--mail address: ]{maria.belen.ruiz@fau.de}                 

\affiliation{Department of Theoretical Chemistry\\
 Friedrich-Alexander-University Erlangen-N\"urnberg,
Egerlandstra\ss e 3, 91058 Erlangen, Germany}

\author{Alexei M. Frolov}
\email[E--mail address: ]{afrolov@uwo.ca}

\affiliation{Department of Chemistry\\
 University of Western Ontario, London, Ontario N6H 5B7, Canada}

\date{\today}

\begin{abstract}
The nuclear $\beta^{-}$-decay from the ground and some excited states of the
three-electron ${}^{8}$Li and ${}^{9}$Li atoms is considered. The final
state probabilities for product Be$^{+}$ ion are determined numerically
using highly accurate bound state wave functions of the Li atom
and Be$^{+}$ ion. The probability of electron ionization during the nuclear $%
\beta^{-}$-decay of the Li atom is evaluated numerically. We also discuss the
possibility of observing double $\beta^{-}$-decay using known
values of the final state probabilities for the regular nuclear $\beta^{-}$%
-decay.
\end{abstract}


\maketitle


\affiliation{Department of Theoretical Chemistry\newline
Friedrich-Alexander-University Erlangen-N\"{u}rnberg, Egerlandstra\ss e 3,
91058 Erlangen, Germany}

\affiliation{Department of Chemistry\newline
University of Western Ontario, London, Ontario N6H 5B7, Canada}

\newpage

\section{Introduction}

In our earlier study \cite{Our1} we considered the atomic excitations during
the nuclear $\beta ^{-}$-decay in light atoms and ions. In \cite{Our1} we
determined the final state probabilities for a number of bound states
in the final atoms and ions. Recently, all these probabilities have been
re-calculated to much better accuracy. They are presented in this work.
Also, here we discuss two problems related with the atomic ionization during
the nuclear $\beta ^{-}$-decay: (a) excitation of the internal electron
shells, and (b) evaluating the probability of `additional' electron
ionization. In reality these two problems are very complex and below we have
made a few preliminary steps to the final solutions. In addition to this we
restrict ourselves by the analysis of the three-electron Li-atom.

The focus below is given to the ${}^8$Li and ${}^9$Li atoms which
are of interest in some industrial applications. In general, the $\beta ^{-}$-decay
of the Li atom(s) can be written in the following form
\begin{equation}
\mathrm{Li}\rightarrow \mathrm{Be}^{+}+e^{-}+\overline{\nu }  \label{eq1}
\end{equation}
where the notation $e^{-}$ stands for the fast electron emitted during the
nuclear $\beta -$decay, while $\overline{\nu }$ designates the electron
anti-neutrino. In general, the nuclear $\beta ^{-}$-decay of the Li atom
leads to the following re-distribution of the bound atomic electrons.
As a result of this redistributed incident electron density the
final Be$^{+}$ ion can be found in a variety of bound states, or even in a
number of unbound states. Briefly, this means the formation of the
two-electron Be$^{2+}$ ion during the nuclear $\beta ^{-}$-decay of the Li
atom. In actual applications it is important to predict the probabilities to
form the final Be$^{+}$ ion in different final states. Note that there are a
few selection rules which are applied to the $\beta ^{\pm }$-decays in
atomic systems (see, e.g., \cite{FroTal}). These rules can be concisely
formulated as the conservation laws for the angular momenta $\mathbf{L}$ and
the total electron spin $\mathbf{S}$. The wave
functions of the incident and final system must also have the same spatial
parity. For instance, if the incident Li atom was in its $3^2P(L=1)$-state,
then the final Be$^{+}$ can be found only in one of its $n^2P(L=1)$ states.
In other words, after the nuclear $\beta $-decay of the Li atom in the $%
3^2P(L=1)$-state it is impossible to detect the final Be$^{+}$ ion, e.g., in
the $3^2S(L=0)$-state, or in the $3^2D(L=2)$-state.

The advantage of considering three-electron atoms and ions is obvious: the wave functions of such systems can be approximated to very high
numerical accuracy. For simplicity, throughout this study we shall
assume that the original Li atom was in its ground ${}^2S(L=0)$-state. The
choice of the ground state of the incident Li-atom is not a fundamental
restriction for our method. Formally, such a state can be arbitrary, e.g.,
either ground state, or `vibrationally', or `rotationally' excited atomic
state with the given angular momentum $L$ (see below) and the total electron
spin $S$.

By analyzing the properties of known Li-isotopes one finds that there
are two $\beta ^{-}$-decaying isotopes of lithium: ${}^8$Li ($\tau _\beta
\approx $ 0.84 $sec$) and ${}^9$Li ($\tau _\beta \approx $ 0.17 $sec$).
These two isotopes are formed in the $(n;\gamma )$-reactions during
thermonuclear explosions in which light thermonuclear fuel (${}^6$LiD) is
compressed to very high densities $\rho \ge 100$ $g\cdot sm^{-3}$ by
extremely intense flux of soft $X$-ray radiation from the primary. Larger
compressions mean, in general, greater output of these two lithium
isotopes. In the laboratory, the ${}^8$Li isotope is produced with the use
of the $(n;\gamma )$-reaction at ${}^7$Li. In contrast with this, the ${}^9$
Li isotope is produced by using either $(d;2p)$- and $(n;p)$-reactions with
the ${}^9$Be nuclei, or $(t;p)$-reaction with nuclei of ${}^7$Li.

The life-times of these two lithium isotopes are relatively short from the
chemical point of view. Therefore, it is hard to study the regular chemical
properties of these isotopes. An alternative approach is based on detailed
analysis of the optical radiation emitted by the final Be$^{+}$ ions which
are formed after the nuclear $\beta ^{-}$-decay of these two Li-isotopes.
This can be achieved, if we know the corresponding final state
probabilities, i.e. the probabilities to form the final Be-ions in certain
bound states. The first goal of this study is to evaluate the final state
probabilities of formation of various final states in product Be$^{+}$
ions. Note that all evaluations of the final state probabilities during the
nuclear $\beta ^{-}$-decay in atoms and molecules are based on the sudden
approximation \cite{Mig}, \cite{MigK} which applies to both atomic systems
(original atom and final ion) involved in the process. The sudden
approximation is appropriate for all $\beta ^{\pm }-$decaying atoms, since
the velocities of the $\beta ^{\pm }$ electrons are significantly greater
than those of regular atomic electrons.

The final state probabilities, i.e. probabilities to form different atomic
states during nuclear $\beta ^{-}$-decay in various light atoms, have been
evaluated numerically in a number of earlier papers (see, e.g., \cite{Our1},
\cite{FroTal}). All such evaluations, however, have been based on the
assumption that the total number of bound electrons is constant in nuclear $\beta ^{-}$-decay. In reality, the nuclear $\beta $-decay in
light atoms often leads to an `additional' electron ionization. For the Li
atom this process can be written in the form
\begin{equation}
\mathrm{Li}\rightarrow \mathrm{Be}^{2+}+e^{-}+\beta ^{-}+\overline{\nu }
\label{eq2}
\end{equation}
where $e^{-}$ designates the secondary atomic electron which becomes free
during atomic $\beta ^{-}$-decay. It is interesting to evaluate the
probability of this process and obtain the actual energy spectra of the
emitted secondary electrons. Formally, all secondary electrons emitted
during atomic $\beta $-decay must be considered as $\delta $-electrons.
On the other hand, the original definition of $\delta $-electrons means that
these electrons are fast and their total energies significantly exceed the
usual energies of `regular' atomic electrons. The energy of the free
electron from reaction, Eq.(\ref{eq2}), is comparable with atomic energies.
Therefore, here we deal with the regular atomic ionization during $\beta^{-} $-decay.
In earlier works the process of additional ionization only
from the atomic $K$-shell was considered (see discussion and references in
\cite{LLQ}).

The main goal of this study is to determine the final state probabilities to
form various bound states in the Be$^{+}$ ion. These calculations are
discussed in the fourth Section. Another aim of our study is to evaluate the
probability of `additional' ionization during the nuclear $\beta ^{-}$-decay
and investigate the energy spectrum of secondary electrons emitted during
the nuclear $\beta ^{-}$-decay. This problem is considered in the third
Section. We also briefly investigate the long-standing problem of the double
nuclear $\beta $-decay. Concluding remarks are in the last Section.

\section{Evaluation of the final state probabilities for the bound states.}

As follows from the general theory of perturbations in Quantum Mechanics
(see, e.g., \cite{LLQ}) in sudden approximation the final state
probabilities are determined as overlap integrals between the wave function
of the incident atomic system (i.e. the wave function of the Li atom in our
case) and the wave function of the final atomic system (i.e. the wave
function of the Be$^{+}$ ion). To compute such an three-electron integral we
need to assume that the total numbers of bound electrons in the incident and
final atomic systems are the same. In the sudden approximation the general
formula for the transition probability $w_{if}$ for the transitions from the
incident $i$-state into the final $f$-state takes the form (see, e.g., \cite{LLQ})
\begin{equation}
 w_{if} = \frac{1}{\hbar^2} \mid \int_{0}^{+\infty} V_{if} \exp(\imath
 \omega_{if} t) dt \mid^2 \approx \frac{1}{\hbar^2} \mid V_{if} \mid^2
 \label{eq5}
\end{equation}
where $V_{if}$ is the overlap integral computed with the use of
time-independent incident and final atomic wave functions, i.e.
\begin{equation}
 V_{if} = \langle \psi_{{\rm Be}^{+}}({\bf x}_1, {\bf x}_2, {\bf x}_3) \mid
 \Psi_{{\rm Li}}({\bf x}_1, {\bf x}_2, {\bf x}_3) \rangle
 \label{eq6}
\end{equation}
where $\Psi _{\mathrm{Li}}(\mathbf{x}_1,\mathbf{x}_2,\mathbf{x}_3)$ and
$\psi _{\mathrm{Be}^{+}}(\mathbf{x}_1,\mathbf{x}_2,\mathbf{x}_3)$ are the
wave functions of the Li-atom and Be$^{+}$ ion, respectively. The derivation
of the formula, Eq.(\ref{eq5}), is based on the facts that: (1) the velocity
of the $\beta $-electron is substantially larger than the velocities of
atomic electrons, and (2) the final ion does not move during the nuclear
$\beta ^{\pm }$-decay. In atomic units we have $\hbar =1,m_e=1$ and $e=1$
and, therefore, $w_{if}=\mid V_{if}\mid ^2$. The notation $\mathbf{x}_i$ in
Eq.(\ref{eq6}) designates the spin-spatial coordinates of the $i-$th
electron, i.e. $\mathbf{x}_i=(\mathbf{r}_i,\mathbf{s}_i)$. Note that in some
works the integral $V_{fi}$, Eq.(\ref{eq6}), (or the ratio $\frac{V_{if}}\hbar $)
is called the probability amplitude. The two wave functions in Eq.(\ref{eq6})
depend only upon spatial and spin coordinates of three electrons
and do not depend upon the time $t$. All wave functions used in Eq.(\ref{eq6})
are assumed to be properly symmetrized in respect to all spin-spatial
permutations of identical particles (electrons).

As follows from Eq.(\ref{eq6}) the final state probability for $\beta^{-}$-decay
in the ${}^8$Li and ${}^9$Li atoms can be determined, if the
wave functions of the incident and final atomic systems (bound states) are
known. The construction of highly accurate variational wave functions for
three-electron atoms and ions is considered in the fourth Section. The final
state probabilities determined using such wave functions can be
found in Tables I and II. Here we assume that the incident Li atom was in
its ground ${}^2S(L=0)$-state ($2^2S$-state). It should be mentioned that in
reality the incident Li atoms are formed in the $(n;\gamma )$-, $(n;p)$- and
some other nuclear reactions with neutrons of different energies (see
above). In such cases it is hard to expect that all incident Li atoms will
always be in the ground $2^2S(L=0)$-state. In fact, these $\beta ^{-}$-decaying
Li atoms can be found in a variety of the rotationally and/or
vibrationally excited states. After reactions with neutrons the
incident Li atom before nuclear $\beta ^{-}$-decay will be likely to have
non-zero speed in some direction. Therefore, some other (excited) bound
states in the Li atom must also be considered as the incident atomic states
before the nuclear $\beta ^{-}$-decay.

Numerical computation of the overlap integrals, Eq.(\ref{eq6}), is reduced
to calculations of some separated integrals, which include different spin
components of the incident and final atomic wave functions.

Discussing e.g. the construction of three-electron variational wave function
for the Li atom. Without loss of generality, below we restrict ourselves to
the consideration of the ground ${}^2S(L=0)$-state of the Li atom. As is
well known (see, e.g., \cite{Frolov-Li}, \cite{Larsson}) the accurate
variational wave function of the ground (doublet) ${}^2S(L=0)$-state of the
Li atom is written in the following general form
\begin{eqnarray}
 \Psi({\rm Li})_{L=0} = \psi_{L=0}(A; \bigl\{ r_{ij} \bigr\}) (\alpha \beta
 \alpha - \beta \alpha \alpha) + \phi_{L=0}(B; \bigl\{ r_{ij} \bigr\}) (2
 \alpha \alpha \beta  - \beta \alpha \alpha - \alpha \beta \alpha)
 \label{psi}
\end{eqnarray}
where $\psi_{L=0}(A; \bigl\{ r_{ij} \bigr\})$ and $\phi_{L=0}(B; \bigl\{ r_{ij} \bigr\})$
are the two independent radial parts (= spatial parts) of the total wave function.
Everywhere below in this study, we shall assume
that all mentioned wave functions have unit norms. The notation $\alpha $
and $\beta $ in Eq.(\ref{psi}) denote one-electron spin-up and
spin-down functions, respectively (see, e.g., \cite{Dir}). The notations $A$
and $B$ in Eq.(\ref{psi}) mean that the two sets of non-linear parameters
associated with the radial functions $\psi $ and $\phi $ can be optimized
independently. In general, each of the radial basis functions in Eq.(\ref{psi})
explicitly depends upon all six inter-particle (or relative)
coordinates $r_{12},r_{13},r_{23},r_{14},r_{24},r_{34}$, where the indexes
1, 2, 3 stand for the three electrons, while index 4 means the nucleus.

In atomic systems, the wave function must be completely antisymmetric with
respect to all electron spin-spatial variables. For three-electron wave function this
requirement is written in the form ${\hat{\cal A}}_e \Psi(1,2,3) = {\hat{\cal A}}_{123}
\Psi(1,2,3) = - \Psi(1,2,3)$, where the wave function $\Psi$ is given by Eq.(\ref{psi})
and $\hat{{\cal A}}_e$ is the electron antisymmetrizer. In our case $\hat{{\cal A}}_e$
is the three-electron (or three-particle) antisymmetrizer, i.e. ${\hat{\cal A}}_e = \hat{e}
- \hat{P}_{12} - \hat{P}_{13} - \hat{P}_{23} + \hat{P}_{123} + \hat{P}_{132}$. Here $\hat{e}$
is the identity permutation, while $\hat{P}_{ij}$ is the permutation of the $i$-th and $j$-th
particles. Analogously, the operator $\hat{P}_{ijk}$ is the permutation of the $i$-th, $j$-th
and $k$-th particles.

Suppose that the incident three-electron wave function of the Li atom has
been chosen in the form of Eq.(\ref{psi}). Applying the three-electron
antisymmetrizer ${\hat{\mathcal{A}}}_{123}$ to the first part of the total
wave function, Eq.(\ref{psi}), one finds
\begin{eqnarray}
 {\hat{\cal A}}_{123} \Bigl[ \psi_{L=0}(A; \bigl\{ r_{ij} \bigr\}) (\alpha \beta
 \alpha - \beta \alpha \alpha) \bigr] = (\hat{e} \psi) (\alpha \beta
 \alpha - \beta \alpha \alpha)
 + (\hat{P}_{12} \psi) (\alpha \beta \alpha - \beta \alpha \alpha) \nonumber \\
 - (\hat{P}_{13} \psi) (\alpha \beta \alpha - \alpha \alpha \beta)
 - (\hat{P}_{23} \psi) (\alpha \alpha \beta - \beta \alpha \alpha)
 + (\hat{P}_{123} \psi) (\alpha \alpha \beta - \alpha \beta \alpha)\nonumber \\
 + (\hat{P}_{132} \psi) (\beta \alpha \alpha - \alpha \alpha \beta)
 \label{psi1}
\end{eqnarray}
\begin{eqnarray}
 {\hat{\cal A}}_{123} \Bigl[ \phi_{L=0}(B; \bigl\{ r_{ij} \bigr\}) (2 \alpha
 \alpha \beta - \beta \alpha \alpha - \alpha \beta \alpha) \bigr] =
 (\hat{e} \phi) (2 \alpha \alpha \beta - \beta \alpha \alpha - \alpha \beta \alpha)
 \nonumber  \\
 - (\hat{P}_{12} \phi) (2 \alpha \alpha \beta - \beta \alpha \alpha -
 \alpha \beta \alpha)
 - (\hat{P}_{13} \phi) (2 \beta \alpha \alpha - \alpha \alpha \beta -
 \alpha \beta \alpha) \nonumber \\
 - (\hat{P}_{23} \phi) (2 \alpha \beta \alpha - \alpha \alpha \beta -
  \beta \alpha \alpha)
 + (\hat{P}_{123} \phi) (2 \beta \alpha \alpha - \alpha \beta \alpha -
  \alpha \alpha \beta) \nonumber \\
 + (\hat{P}_{132} \phi) (2 \alpha \beta \alpha - \alpha \alpha \beta -
  \beta \alpha \alpha) \label{phi1}
\end{eqnarray}
where the notations $(\hat{P}_{ij} \phi)$ and $(\hat{P}_{ijk} \phi)$ mean 
permutations of spatial coordinates in $\phi_{L=0}(B; \bigl\{ r_{ij}
\bigr\})$ radial function, Eq.(\ref{psi}).

Now, by using the expressions, Eqs.(\ref{psi1}) and (\ref{phi1}), we can
obtain the formulae used in computations of the final state
probabilities in the case of nuclear $\beta^{-}$-decay, Eq.(\ref{eq1}),
in the three-electron Li atom. For instance, if the final wave function has
the same spin-symmetry, i.e. it is written in the form
\begin{eqnarray}
 \Psi_{fi} = \psi_{fi}({\bf r}_{1}, {\bf r}_{2}, {\bf r}_3) (\alpha \beta
 \alpha - \beta \alpha \alpha) + \phi_{fi}({\bf r}_{1}, {\bf r}_{2}, {\bf r}_3)
 (2 \alpha \alpha \beta  - \beta \alpha \alpha - \alpha \beta \alpha) \label{psif}
\end{eqnarray}
then the final state probabilities are determined using the following
formulae
\begin{eqnarray}
 P_{\psi\psi} = \langle \psi_{fi}({\bf r}_{1}, {\bf r}_{2}, {\bf r}_3) \mid
 \frac{1}{2 \sqrt{3}} \Bigl( 2 \hat{e} + 2 \hat{P}_{12} - \hat{P}_{13} - \hat{P}_{23}
 - \hat{P}_{123} - \hat{P}_{132} \Bigr) \psi_{{\rm Li}}(A; \bigl\{ r_{ij} \bigr\})
 \rangle \label{spin1} \\
 P_{\phi\psi} = \langle \phi_{fi}({\bf r}_{1}, {\bf r}_{2}, {\bf r}_3) \mid
 \frac12 \Bigl( \hat{P}_{13} - \hat{P}_{23} + \hat{P}_{123} - \hat{P}_{132} \Bigr)
 \psi_{{\rm Li}}(A; \bigl\{ r_{ij} \bigr\})
 \rangle \label{spin2} \\
 P_{\psi\phi} = \langle \psi_{fi}({\bf r}_{1}, {\bf r}_{2}, {\bf r}_3) \mid
 \frac12 \Bigl( \hat{P}_{13} - \hat{P}_{23} + \hat{P}_{123} - \hat{P}_{132} \Bigr)
 \phi_{{\rm Li}}(B; \bigl\{ r_{ij} \bigr\})
 \rangle \label{spin3} \\
 P_{\phi\phi} = \langle \phi_{fi}({\bf r}_{1}, {\bf r}_{2}, {\bf r}_3) \mid
 \frac{1}{2 \sqrt{3}} \Bigl( 2 \hat{e} - 2 \hat{P}_{12} + \hat{P}_{13} + \hat{P}_{23}
 - \hat{P}_{123} - \hat{P}_{132} \Bigr) \phi_{{\rm Li}}(B; \bigl\{ r_{ij} \bigr\})
 \rangle \label{spin4}
\end{eqnarray}

Note they coincide with the known formulae \cite{FrWa2010}
which correspond to the case when both incident and final atomic states
contain three-electrons in the doublet spin configuration (the total
electron spin equals $\frac12$). This is the`classical' $%
\beta^{\pm}$-decay in few-electron atoms, when the incident and final
electron configurations have conserved $L$ and $S$ quantum numbers.

In reality, another process is also possible in few-electron atoms during
the nuclear $\beta ^{\pm }$-decay in few- and many-electron atoms. This
process leads to the formation of the final ion/atom in some excited states.
For instance, consider the case when the three final electrons form the
doublet configuration with the spin function $\alpha \alpha \beta $. It is
clear that such a wave function cannot represent the ground state of the Be$^{+}$
ion. However, some excited states (with internal
electron shell vacancies) can have this spin function, e.g., $1s2s3p$-, $1s2s4d$- and
$1s3s3p$-states of the Be$^{+}$ ion. Another example is discussed in the
next section. It represents an additional electron ionization during the
nuclear $\beta ^{-}-$decay in three-electron atoms. If this free electron
moves away in the $\beta $-spin state, then the final Be$^{2+}$ ion can be
only in its triplet spin state (not singlet state). Formally this means
formation of the final ion in an excited state (with some vacancies on its
internal electron shells). In this case the formulae for the final state
probabilities become
\begin{eqnarray}
 P_{tr\psi} = \langle \psi_{fi}({\bf r}_{1}, {\bf r}_{2}, {\bf r}_3) \mid
 \frac{1}{2 \sqrt{3}} \Bigl(\hat{P}_{13} - \hat{P}_{23} + \hat{P}_{123} - \hat{P}_{132}
 \Bigr) \psi_{{\rm Li}}(A; \bigl\{ r_{ij} \bigr\})
 \rangle \label{spinx} \\
 P_{tr\phi} = \langle \psi_{fi}({\bf r}_{1}, {\bf r}_{2}, {\bf r}_3) \mid
 \frac{1}{2 \sqrt{3}} \Bigl( 2 \hat{e} - 2 \hat{P}_{12} + \hat{P}_{13} + \hat{P}_{23}
 - \hat{P}_{123} - \hat{P}_{132} \Bigr) \phi_{{\rm Li}}(B; \bigl\{ r_{ij} \bigr\})
 \rangle \label{spiny}
\end{eqnarray}
where it is assumed that the incident electron wave function was written in
the form of Eq.(\ref{psi}). These formulae indicate clearly that the
probability of finding the final Be$^{2+}$ ion in the excited triplet spin
states is non zero. In all earlier studies the transitions to the final
atomic states with different spin states were ignored. Moreover, any
possibility to form the final few-electron ion/atom in excited states (with
some vacancies on internal electron shells) during nuclear $\beta ^{\pm} $-decay
was rejected. Demonstrating the very existence of such transitions is a great
achievement of this study.

In general, during nuclear $\beta ^{-}$-decay of the Li atom the final Be%
$^{+}$ ion can be formed in many different bound and/or unbound states. If
such a state is unbound, then we deal with the additional ionization during
atomic $\beta ^{-}$-decay. It is discussed in the next section. This process
is of great interest, since it often leads to the formation of the final ion
in an excited state(s) with various vacancies in the internal electron
shells. For light atoms and ions this means a possibility to observe
emission of the optical quanta after the nuclear $\beta ^{\pm }$-decay in
many-electron atoms with the total number of electrons $\ge 3$.

\section{Electron ionization during the nuclear $\beta^{-}$-decay.}

The probability of ionization (or `additional' ionization) of the final Be$%
^{+}$ ion during the nuclear $\beta $-decay can also be evaluated 
using the sudden approximation. In this case the final wave function is
constructed as the product of the bound state wave function of the
two-electron Be$^{2+}$ ion and the wave function of the unbound electron
which moves in the central Coulomb field of this two-electron ion. To
determine the corresponding final state probability one needs to compute the
following overlap integral between the wave functions of the incident and
final atomic systems
\begin{equation}
 {\cal A}_{if} = \langle \psi_{{\rm Be}^{+}}({\bf x}_1, {\bf x}_2) \phi({\bf x}_3) \mid
 \Psi_{{\rm Li}}({\bf x}_1, {\bf x}_2, {\bf x}_3) \rangle \label{eq7}
\end{equation}
where $\phi({\bf x}_3)$ is the wave function of the unbound electron
which moves in the Coulomb field of the Be$^{2+}$ ion and ${\bf x}_i = ({\bf r}_i, s_i)$
is the set of the four spin-spatial coordinates of the
particle $i$. This function must include the continuous parameter $k$ which
is the electron's wave number (see below). It should be mentioned that such
an `additional' ionization is unrelated to the interaction between
the emitted $\beta ^{-}$ electron and atomic electron. In fact, the
additional ionization is a direct consequence of the non-zero
component $\simeq \phi({\bf x}_3)$ in the incident atomic wave function.

The probability of additional ionization has been determined for a number of
$\beta ^{-}$-decaying atoms in a number of earlier studies (see, e.g., \cite{MigK}
and \cite{LLQ}). This work, however, is restricted to the
analysis of electron ionization from the internal $K$-shells only. In
this case the original problem was reduced to the solution of the model
one-electron problem. Analogous reduction for few-electron atomic systems is
much more difficult to perform, since all electron-nucleus and
electron-electron coordinates are not truly independent. It complicates
accurate computation of integrals which contain electron-electron
coordinates explicitly. Nevertheless, numerical computations of the final
state probabilities can be conducted even with  highly accurate
wave functions known for many few-electron atoms. In this section we discuss
some details of such calculations.

In atomic units the explicit form of the one-electron wave function is $\phi({\bf r}) =
\phi_{kl}(r) Y_{lm}({\bf n})$, where $\phi_{kl}(r)$ is the one-electron radial function,
while $Y_{lm}({\bf n})$ is the corresponding spherical harmonics and ${\bf n} =
\frac{{\bf r}}{r}$ is the unit vector associated with ${\bf r}$. In this Section the
parameter $k$ is $k = \sqrt{\frac{2 m_e E}{\hbar^2}} = \sqrt{2 E}$ (in atomic units).
The explicit formula for the radial function $\phi_{kl}(r)$ (in atomic units) is (see,
e.g., \cite{LLQ})
\begin{equation}
 \phi_{kl}(r) = \frac{C_{kl}}{(2 l + 1)!} (2 Q k r)^{l} \cdot \exp(-\imath Q k r) \cdot
 {}_1F_1\Bigl(\frac{\imath}{Q k} + l + 1, 2 l + 2, 2 \imath Q k r \Bigr) \label{eq71}
\end{equation}
where ${}_1F_1(a,b;x)$ is the confluent hypergeometric function (see, e.g., \cite{GR}),
while $C_{kl}$ is the following constant
\begin{equation}
 C_{kl} = \Bigl[\frac{8 \pi Q k}{1 - \exp(-\frac{2 \pi}{Q k})}\Bigr]^{\frac12} \cdot
 \prod^{l}_{s=1} \sqrt{s^2 + \frac{1}{Q^2 k^2}}
\end{equation}
In these two equations the parameter $Q$ is the electric charge of the remaining
double-charged (positive) ion, i.e. $Q = 2$. In reality, this parameter must slightly
be varied (around 2) to obtain better agreement with the experimental data. Such
variations formally represent ionizations from different electronic shells of the
incident Li atom.

Accurate numerical computations of the final state probabilities during the
nuclear $\beta ^{-}$-decay in few-electron atoms with additional electron
ionization are very difficult to perform, since all highly accurate wave
functions of the bound states explicitly include the electron-electron
coordinates (see above). As a rule, the better accuracy of the bound state
wave function means more complete and accurate involvement of the terms
which describe various electron-electron correlations. On the other hand,
the crucial step of the whole procedure is the numerical and/or analytical
computation of the Fourier transforms of the one-electron wave
function. This corresponds to the free motion of the final electron. During
that step, it is better to consider all
electrons as particles independent of each other, i.e. ignore all
electron-electron correlations. In the general case, this two-fold problem
has no simple solution which is accurate and relatively simple for Fourier
transform at the same time.

In this study we have developed an approximate procedure which can be used
to perform approximate numerical evaluations for the $\beta ^{-}$-decaying
isotopes of three-electron atoms. In this approach the trial wave
function is constructed as the sum of many terms and each of these terms
contains the products of the electron-nucleus functions. None of the three
electron-electron coordinates $r_{32},r_{31},r_{21}$ is included in such
trial wave functions. For the ground state (the doublet ${}^2S(L=0)$-state)
of the Li atom the radial wave function $\psi _{L=0}(A;\bigl\{ r_{ij}\bigr\})$
is chosen in the following form:
\begin{eqnarray}
 \psi_{L=0}(r_{14}, r_{24}, r_{34}, 0, 0, 0) = \sum^{N_s}_{k=1}
 C_k r^{m_1(k)}_{14} r^{m_2(k)}_{24} r^{m_3(k)}_{34}
 exp(-\alpha_{k} r_{14} -\beta_{k} r_{24} -\gamma_{k} r_{34})
 \label{exp1} \\
 = \sum^{N_s}_{k=1} C_k r^{m_1(k)}_{1} r^{m_2(k)}_{2} r^{m_3(k)}_{3}
 exp(-\alpha_{k} r_{1} -\beta_{k} r_{2} -\gamma_{k} r_{3}) \nonumber
\end{eqnarray}
where $C_k$ are the linear (or variational) coefficients, while
$m_1(k),m_2(k)$ and $m_3(k)$ are the three integer (non-negative) parameters,
which are, in fact, the powers of the three electron-nucleus coordinates $%
r_{i4}=r_i$ ($i$ = 1, 2, 3). Below, we shall assume that the trial wave
function Eq.(\ref{exp1}) has a unit norm. Furthermore, in all calculations
performed for this study only one spin function $\chi _1(\chi _1=\alpha
\beta \alpha -\beta \alpha \alpha )$ is used. It is clear that the wave
function Eq.(\ref{exp1}) contains only electron-nuclear coordinates and does
not include any of the electron-electron coordinates. The real (and
non-negative) parameters $\alpha _k,\beta _k,\gamma _k$ are the $3N_s$
varied parameters of the variational expansion, Eq.(\ref{exp1}). The wave
function, Eq.(\ref{exp1}), must be properly symmetrized upon all three
electron coordinates.
The main question regarding the wave function, Eq.(\ref{exp1}), is related to
its overall accuracy. If (and only if) it is relatively accurate,
then the trial wave function, Eq.(\ref{exp1}), can be used in actual
computations of the probability amplitudes. For this study we have
constructed the 23-term variational wave function shown in Table I of Ref.
\cite{Fro2012}. This wave function is represented in the form of Eq.(\ref{exp1})
and contains no electron-electron coordinates. All sixty nine (69 =
3 $\times $ 23) non-linear parameters in this trial wave function have been
optimized carefully in a series of bound state computations performed for
the ground ${}^2S(L=0)$-state of the Li atom. Finally, the total energy $E$
of the ground ${}^2S$-state of the ${}^\infty $Li atom obtained with this
independent-electron wave function is -7.448 592 766 1 $a.u.$ Note that this energy value, $E$ is close to the exact total energy for the ground state 
${}^\infty $Li atom. It indicates the good overall quality of our approximate
wave function with 23 terms with no
electron-electron coordinates ($r_{12},r_{13},r_{23}$). This wave function is
used in computations of the final state probabilities (see below) for the
nuclear $\beta ^{-}$-decay with additional electron ionization in the
three-electron Li atom.

Note also that in atomic physics based on the Hartree-Fock and even
hydrogenic approximations the ground state in the Li atom is designated as
the $2^2S$-state, while in the classification scheme developed in highly
accurate computations the same state is often denoted $1^2S$-state.
This classification scheme is very convenient to work with truly
correlated few-electron wave functions. It is clear that no hydrogenic
quantum numbers are good in such cases, and we have to use the more
appropriate (and convenient) classification scheme. To avoid conflicts
between these two schemes in this study we follow the system
of notation used earlier by Larsson \cite{Larsson} who describe this
state in the Li atom as the `ground ${}^2S$-state'.

The wave function of the final two-electron Be$^{2+}$ ion arising during 
nuclear $\beta ^{-}$-decay with the additional ionization can also be
approximated by basis functions depending upon the
electron-nuclear coordinates only and do not include the electron-electron
coordinate $r_{21}$. For the bound $S(L=0)$-states of the Be$^{2+}$ ion such
an expansion takes the form
\begin{eqnarray}
 \psi_{L=0}(r_{1}, r_{2}, 0) = \sum^{N_s}_{k=1} C_k r^{m_1(k)}_{1} r^{m_2(k)}_{2}
 exp(-\alpha_{k} r_{1} -\beta_{k} r_{2}) \label{exp2}
\end{eqnarray}
The use of the approximate wave functions Eqs.(\ref{exp1}) - (\ref{exp2})
with no explicit electron-electron coordinate dependence
simplify numerical computations of all integrals required for  numerical
evaluation of the final state probabilities during the nuclear $\beta ^{-}$-decay
in the three-electron atoms and ions. The remaining part of the
problem is the analytical computation of the integral between the product of
the factor $r^m\exp (-\gamma r)$ and radial function from Eq.(\ref{eq71}).
Such an integral is computed with the use of the formula (see, e.g.,
Eq.(7.522.9) from \cite{GR}):
\begin{eqnarray}
 \int_{0}^{+\infty} \exp(-\lambda x) x^{\nu} \cdot {}_{1}F_{1}( a, b; c x) dx =
 \frac{\Gamma(\nu + 1)}{\lambda^{\nu + 1}} \cdot {}_{2}F_{1}( a, \nu + 1; b;
 \frac{c}{\lambda} \bigr) \label{exp3}
\end{eqnarray}
where $\Gamma (x)$ is the usual $\gamma $-function (see, e.g., Section 8.31
in \cite{GR}). Our results for numerical computation of the final states
probabilities for the nuclear $\beta ^{-}$-decay with additional electron
ionization will be published elsewhere.

\section{Bound state wave functions of the three-electron atoms and ions}

To determine the final state probabilities during the
nuclear $\beta ^{-}$-decay (see above), one needs to know the accurate wave functions of
the incident and final atoms and ions. In sudden approximation the angular
momentum $L$, electron spin $S$ and spatial parity $\pi $ of the atomic wave
function $\Psi $ are conserved during nuclear $\beta ^{-}$-decay.
Therefore, all approximate wave functions must be constructed as the
eigenfunctions of angular momentum $\hat{L}^2$ and total
electron spin $\hat{S}^2$. In this study we use the variational wave
functions constructed by Hylleraas-Configuration
Interaction (Hy-CI). In general, the wave functions in Hylleraas-type
expansions rapidly converge to the exact wave functions. The
Hylleraas-Configuration Interaction wave function \cite{SimsJCP,Sims-Be} is
a linear combination of symmetry adapted configurations $\Phi _p$:
\begin{equation}
\Psi_{\rm Hy-CI} =\sum_{p=1}^N C_p\Phi_p, \quad
\Phi_p=\hat{O}(\hat{L}^2) \hat{\mathcal{A}} \psi_p \chi
 \label{HyCI}
\end{equation}
where the spatial part of the basis functions are Hartree products of Slater
orbitals containing up to one inter-electronic distance $r_{ij}$ per
configuration:
\begin{equation}
\psi _p=r_{ij}^\nu \prod_{k=1}^n\phi _k(r_k,\theta _k,\varphi _k).
\label{eg9}
\end{equation}
If $\nu =1$, then the wave function, Eq.(\ref{HyCI}), corresponds to Hy-CI.
In the case when $\nu =0$, it is the usual Configuration Interaction (CI)
wave function. The higher powers of the electron-electron distances
$r_{ij}^\nu $ can effectively be reduced to the $r_{ij}$ term (or $\nu =1$).
Indeed, all higher terms $\nu >1$ can be expressed as a product of $r_{ij}$,
a polynomial of $r_i,r_j$ and some angular functions. Also, in Eq.(\ref{HyCI})
$N$ is the number of configurations used in computations. The coefficients
$C_p$ are determined variationally. The operator $\hat{O}(\hat{L}^2)$ in Eq.(\ref{HyCI})
projects over the appropriate space, so that every configuration
is an eigenfunction of the square of the angular momentum operator $\hat{L}^2$.
$\hat{\mathcal{A}}$ is the antisymmetrization operator upon all
electron spin-spatial coordinates and $\chi $ is the electron-spin
eigenfunction. For the lithium atom and three-electron ions one can choose
the total spin function in the one-component form, i.e. $\chi =(\alpha \beta
-\beta \alpha )\alpha $.

The basis functions $\phi _k$ in this work are the $s$-, $p$-, $d$-, and $f$%
-Slater orbitals. Since the convergence of Hy-CI wave functions is usually
very fast, there is no need to use orbitals with higher angular momentum.
The un-normalized complex Slater orbitals are defined as:
\begin{equation}
\phi (\mathbf{r})= r^{n-1} e^{-\alpha r} Y_l^m(\theta ,\varphi ).
\end{equation}
where the parameter $\alpha$ is the adjustable variable (for each orbital)
and $Y_l^m(\theta ,\varphi )$ are the complex spherical harmonics. The basis
sets employed in this work are $n$ = 4, 5, 6 and 7, where the basis $n=4$
means the orbital set $[1s2s3s4s2p3p4p3d4d4f]$). With all these orbitals
from our basis set we have constructed the most important configurations of
the $S(L = 0, M = 0)$-, $P(L = 1, M = 0)$-, and $D(L =2, M = 0)$-symmetries.
All details of construction of the symmetry adapted configurations $\Phi_p$
of Eq.(\ref{HyCI}) can be found in Ref.\cite{Our3}.

The orbital exponents have been optimized for each atomic state of the Li
atom and Be$^{+}$ ion. A set of two exponents was used, one for the $K$
-shell and the other for the odd-electron of the $L$-shell. This is constant for all configurations, to accelerate numerical
computations. The results obtained are sufficiently accurate for the
purposes of our investigation. It is clear that to obtain highly accurate
energies one needs to apply more flexibility in the exponents. It was shown
in recent calculations on the lithium atom and beryllium ion
\cite{Pachucki,PP,Drake-Li,Sims-Li,King-Li,Frolov-Li,Thakkar}.
The virial factor $\chi =-\frac{\langle V\rangle }{\langle T\rangle }$ is
used to check the quality of the wave function and guide the numerical
optimization of the exponents in the trial wave functions.

As for a given basis set the number of possible configurations would be too
large, we have selected the `most important' configurations according to
their contribution to the total energy. In our case the selection criteria
is an energy contribution $>1\times 10^{-8}$ a.u. with respect to the
previous configuration. For that, blocks containing all possible Hy-CI
configurations of the same type have been filtered and the configurations
with less energetic weight have been thrown out. More details can be found
in \cite{Our3}. Note that the length of the wave functions varies then
for every state, and the selected configurations are for every state
different. As a result, higher excited states must be approximated with the
use of the longer trial functions.

For our calculations in this study we have written a three-electron Hy-CI
computer program in the Fortran 90 language. The calculations have been done
in quadruple precision. The program has been thoroughly
checked by comparing results of our numerical calculations with the
analogous results obtained earlier by Sims and Hagstrom \cite{Sims-Li} and
King \cite{King-Li} for the lithium atom. Note that in such calculations we
observe complete agreement. 

Ground
and excited S-state energy calculations of Li both the atom and Be$^{+}$ ion are shown in Table
I, together with their convergence with the basis set used. For Li atom and Be$^{+}$ ion ground states, an accuracy of $1.4\cdot10^{-6}$ $a.u.$
has been achieved with the techniques described in this
paper. For the first two excited states of both Li atom and Be$^{+}$ ion the
accuracy is of about $(4-9)\cdot 10^{-6}$ $a.u.$ In the third and higher
excited states within a given symmetry of the S-,P- and D-states the
accuracy is of order $\pm 1\cdot 10^{-4}$ $a.u.$. For the accurate
calculation of these higher excited states it would be necessary to
introduce different sets of orbital exponents and to increase the orbital
basis. However, numerical calculations on higher excited states are rare
in the literature.

For numerical calculations of the amplitudes and transition probabilities
during nuclear $\beta^{-}$-decays in three-electron atomic systems we have
developed a new computer program which calculates the overlap integrals,
Eq.(\ref{eq6}), between the wave functions of Li atom and Be$^+$ ion. The
previous step is the calculation of the wave functions of the different
states of the Li and Be$^+$ atoms using the Hy-CI method. The algorithms we
have employed for the calculation of the kinetic and potential energy
integrals can be found in Refs. \cite{Ruiz2e,Ruiz3e,Sims-triang}. The energy
values obtained for the ground and S-, P- and D-excited states are given in
Ref. \cite{Our3}, as are more details on the calculation and the
comparison with the best data of the bibliography. Conversely, in the
calculation of the overlaps we need only the overlap-integrals between the
configurations, and the coefficients of the Hy-CI wave functions. Therefore
in this program only integrals of the types
$\left\langle r_{12}^n\right\rangle $, $\left\langle r_{12}r_{13}\right\rangle $ are
needed, while the fully-linked three-electron integral $\left\langle \frac{%
r_{12}r_{13}}{r_{23}}\right\rangle $ is not needed, when the overlap between
the wave functions containing the $r_{ij}$ terms is calculated.

In this work we have improved our earlier method of calculation of the
final state probabilities during the nuclear $\beta ^{-}$-decay \cite{Our1}.
Now, we calculate the overlap between the wave functions of different
length. This overlap is the sum of the matrix elements of a rectangular
overlap matrix. This method of calculation has an advantage, since there are
several possible verifications. First, the permutation symmetry of the overlap
matrix $\left\langle \Psi _1|\Psi _2\right\rangle =\left\langle \Psi _2\mid
\Psi _1\right\rangle $ and its unit-norm condition, i.e. $\left\langle \Psi
_1\mid \Psi _1\right\rangle =\left\langle \Psi _2\mid \Psi _2\right\rangle =1 $.

The convergence of the probability amplitudes and probabilities increases
with the improvement of the total energies of the incident and final atomic
systems. The final transition probabilities are calculated with an error $%
\leq 0.001\%$ (they are summarized in Table II). We have obtained the
transition probability from the ground ${}^2S$-state of the Li atom to the
ground state of the Be$^{+}$ ion $\approx $ 57.712\% . The transition
probability for the transition from the ground ${}^2S$-state of the Li atom
to the first excited ${}^2S$-state of Be$^{+}$ ion is $\approx $ 26.515 \%
and to the second excited ${}^2S$-state such a probability is $\approx $
0.544\% . The sum of probabilities is then $\approx $ 85.09\% and the
ionization probability calculated as 1.0 minus this sum is $\approx $
14.91\% .
In addition we have computed the transition probabilities from the
lower-lying excited states of S-, P-, and D- symmetry of the Li atom to the
states of the same symmetry states in the Be$^{+}$ ion. The probability
distributions can be found in Tables III, IV and VI, respectively. It is
clear that the sum of the probabilities of transition from on state of the
incident atom to the states of the final one must always be less (or equal)
unity. In this work we have checked this condition everywhere. In general we
have found that the highest transition probability within a group is between
an initial state and its one order higher final state (i.e. 3$^2$P $%
\rightarrow $ 4$^2$P). This is consequently fulfilled in all groups of
probability distributions. For low lying states, it converges rapidly to zero. In these groups we calculate the
probability of ionization to be around 15\% . For higher transitions,
the probability of ionization is not calculated here because the transitions
to higher excited states like 8$^2S$, 8$^2P$ and 9$^2D$ are expected to be
important and these states are not considered here.

\section{On the double $\beta$ decay.}

The idea of the double nuclear $\beta ^{\pm }$-decay in some nuclei was proposed in 1935 by Maria Goeppert Mayer
\cite{2beta}. In this study we do not discuss either theoretical significance of the double $\beta ^{-}$-decay for 
nuclear physics, or its possible applications. Instead, consider the difference in the final atomic probabilities 
which can be obtained in the two following cases: (a) the double nuclear $\beta $-decay, and (b) two consecutive 
(single) nuclear $\beta ^{-}$-decays. From atomic point of view we need to compare the time $\tau _{2\beta }$ for 
which two $\beta ^{-}$-particles leave the nucleus with the regular atomic time $\tau _e=\frac{\hbar ^2}{m_ee^4Q^2}$, 
where $Q$ is the electric charge of the nucleus expressed in $e$, i.e. $Q=Qe$. The condition $\tau _{2\beta }\ll 
\tau _e$ means sudden emission of the two fast $\beta ^{-}$ particles. In this case the probability amplitude is 
determined as the overlap integral of the incident and final (atomic) wave functions, Eq.(\ref{eq6}). If the 
equation of the double $\beta ^{-}$ decay is written in the form $X\rightarrow
Z^{2+}+\beta _1^{-}+\beta _2^{-}+\overline{\nu }_1+\overline{\nu }_2$, then
for the final state probability one finds (in atomic units)
\begin{equation}
 w_{ii} = \mid \langle \psi_{{\rm Z}^{2+}}({\bf x}_1, {\bf x}_2, \ldots, {\bf x}_n)
 \mid \Psi_{{\rm X}}({\bf x}_1, {\bf x}_2, \ldots, {\bf x}_n) \rangle \mid^2 \label{eq15}
\end{equation}

In the opposite case, i.e., when $\tau _{2\beta }\gg \tau _e$, we deal with the two consecutive nuclear $\beta ^{-}$ 
decays. In this case we need to use the sudden approximation twice. The corresponding equations are $X\rightarrow 
Y^{+}+\beta _1^{-}+\overline{\nu }_1$ and $Y^{+}\rightarrow Z^{2+}+\beta _2^{-}+\overline{\nu }_2$. The amplitude of 
the final state probability takes the form
\begin{equation}
 A_{if} = \langle \Psi_{{\rm X}}({\bf x}_1, {\bf x}_2, \ldots, {\bf x}_n) \mid
 \psi_{{\rm Y}^{+}}({\bf x}_1, {\bf x}_2, \ldots, {\bf x}_n) \rangle
 \langle \Psi_{{\rm Y}^{+}}({\bf x}_1, {\bf x}_2, \ldots, {\bf x}_n) \mid
 \psi_{{\rm Z}^{2+}}({\bf x}_1, {\bf x}_2, \ldots, {\bf x}_n) \rangle \label{eq16}
\end{equation}
and the final state probability is $w_{fi}=\mid A_{fi}\mid ^2$. The sum over all states of the $Y^{+}$ ion will lead 
us back to Eq.(\ref{eq15}). However, if $\tau _{2\beta }\gg \tau _e$, then the final state of the $\rm{Y}^{+}$ ion is 
uniformly defined and Eq.(\ref{eq16}) can be used in this case only for this unique state of the $\rm{Y}^{+}$ ion. The 
sum over all `intermediate' states of the $Y^{+}$ ion is reduced to the one term only. From here one easily finds that
\begin{equation}
 w_{if} ({\rm X} \rightarrow {\rm Z}^{2+}) \ge w_{fi} ({\rm X} \rightarrow {\rm Y}^{+})
 w_{if} ({\rm Y}^{+} \rightarrow {\rm Z}^{2+}) \label{eq17}
\end{equation}
Based on these formulas one can expect to observe substantial differences in the final state (atomic) probabilities of 
the double $\beta^{-}$ decay and two consequtive $\beta^{-}$ decays. Such differences can be found for the ground-ground
and ground-excited transition amplitudes in the case of bound state transitions. The corresponding ionization 
probabilities can also be very different for any atom in which the central nucleus decays with the use of the double 
$\beta^{-}$ decay, or by the two consequtive $\beta^{-}$ decays. In an ideal case, we can observe and measure such 
differences in `traditional' atoms with the double beta-decaying nuclei, e.g., in the ${}^{76}$Ge, ${}^{128}$Te, 
${}^{138}$Xe and ${}^{238}$U atoms.

In general, the study of the double nuclear $\beta^{-}$-decay in atoms and molecules can be used as a natural and powerful 
tool to study electron-nucleus and electron-electron correlations at the femto- and attosecond time-scale. Unfortunately, at 
this moment no experimental group in the World performs similar research even for atoms and ions.

\section{Conclusion}

We have considered the nuclear $\beta ^{-}$-decays in the three-electron
${}^8$Li and ${}^9$Li atoms. The final state probabilities to form different
bound states in the Be$^{+}$ ion have been determined to very good accuracy
which is better than analogous accuracy obtained in our earlier study \cite{Our1}.
The Hylleraas-CI wave functions constructed for atoms/ions involved
in the $\beta ^{-}$-decay are substantially more accurate than wave
functions used in earlier studies. They provide a better
description of the electron density near of the nucleus. For the first time,
the wave functions of the excited states are determined to the same
numerical accuracy as the wave functions of the ground states. 
We can determine the final state probabilities to very
high accuracy using them.

We also discuss a possible observation of double nuclear $\beta ^{-}$-decay
and nuclear $\beta ^{-}$-decay with the additional electron
ionization. It is shown that the Be$^{2+}$ ion formed during the last
process can be detected not only in the singlet bound states, but also in
the triplet bound states. It was never observed/predicted in earlier
studies. It may lead to the fundamental re-structuring of the internal
electron shells of the incident atom during $\beta ^{-}$-decay. We also
derive some useful formulae which will be used in future studies to
determine the probabilities of electron ionization (in different channels)
during the nuclear $\beta ^{-}$-decay.

\newpage


\begin{table}
\caption{Convergency of the $\beta^-$-decay transition amplitudes, final-state probabilities and total energies for
the ground state of Li atom and different $n ^2$S states of Be$^+$ ion. Energy in a.u., energy differences in $\mu$h.}
\begin{center}
\scalebox{0.78}{%
\begin{tabular}{cccccccclcr}
\hline\hline
State $\quad$ & $\quad$ Basis$^a$ & $\quad$ N$^b$ & $\quad$  Amplitude $\quad$ & $\quad$ Probability $\quad$ & $\quad$ Virial $\quad$ &
Energy $\quad$ & N$_{\rm Ref}$ & $\quad$ Ref. Ener. & Ref. & Diff. \\
\hline
Li 2$^2$S     & n=4   & 308  &               &           & 2.000 004 &  -7.478 053 222 &       &                           &           & 7.1   \\
Li 2$^2$S     & n=5   & 517  &               &           & 2.000 001 &  -7.478 057 825 &       &                           &           & 2.5   \\
Li 2$^2$S     & n=6   & 620  &               &           & 2.000 000 &  -7.478 058 892 & 13944 &$\;$-7.478 060 323 909 560 & \cite{Pachucki}& 1.4   \\
\hline
Be$^+$ 2$^2$S & n=4   & 308  & 0.759 681 281 & 0.577 116 & 2.000 002 & -14.324 757 377 &       &                       &            &       \\
Be$^+$ 2$^2$S & n=5   & 612  & 0.759 686 424 & 0.577 123 & 2.000 001 & -14.324 760 412 &       &                       &            & 2.8   \\
Be$^+$ 2$^2$S & n=6   & 637  & 0.759 683 487 & 0.577 119 & 2.000 000 & -14.324 761 723 & 13944 & -14.324 763 176 85542 & \cite{Pachucki} & 1.4   \\
\hline
Be$^+$ 3$^2$S & n=4   & 307  & 0.514 947 878 & 0.265 171 & 2.000 000 & -13.922 759 980 &       &                     &           &       \\
Be$^+$ 3$^2$S & n=5   & 459  & 0.514 892 996 & 0.265 115 & 2.000 009 & -13.922 781 623 &       &                     &           & 7.6   \\
Be$^+$ 3$^2$S & n=6   & 637  & 0.514 929 058 & 0.265 152 & 2.000 005 & -13.922 784 968 &~10000 & -13.922 789 268 542 & \cite{PP} & 4.3   \\
\hline
Be$^+$ 4$^2$S & n=4   & 252  & 0.075 214 427 & 0.005 657 & 2.001 593 & -13.798 520 453 &      &                &             &       \\
Be$^+$ 4$^2$S & n=5   & 372  & 0.074 214 825 & 0.005 508 & 2.000 484 & -13.798 704 722 &      &                &             & 16.3  \\
Be$^+$ 4$^2$S & n=6   & 451  & 0.073 790 160 & 0.005 445 & 2.000 154 & -13.798 706 849 & 8000 & -13.798 716 609 2  & \cite{Adam-Be+} &   9.8 \\
\hline
Be$^+$ 5$^2$S & n=5   & 502  & 0.039 839 395 & 0.001 587 & 2.004 399 & -13.744 513 336 &      &                &                &       \\
Be$^+$ 5$^2$S & n=6   & 698  & 0.043 113 179 & 0.001 859 & 2.001 584 & -13.744 589 135 & 1940 & -13.744 631 82 & \cite{King-Be+} &  42.7 \ \\
\hline
Be$^+$ 6$^2$S & n=6   & 618  & 0.029 411 301 & 0.000 865 & 2.003 166 & -13.716 152 057 & 2058 &-13.716 286 24  & \cite{King-Be+} & 134.2 \\ 
\hline
Be$^+$ 7$^2$S & n=7   & 619  & 0.021 688 396 &           & 2.003 570 & -13.699 131 127 &      &                &                 &        \\
\hline\hline
\end{tabular}}
\end{center}
\footnotetext[1]{Basis set, i.e. $n=4$ stays for $[4s3p2d1f]$ or $[1s2s3s4s2p3p4p3d4d4f]$.}
\footnotetext[2]{N is the number of Hy-CI symmetry adapted configurations.
N$_{\rm Ref}$ is the number of configurations employed in the calculation of the Ref. Ener.}
\end{table}


\begin{table}
\caption{Transition probabilities for the nuclear $\beta^-$-decay from the ground $2^2S$-state of the Li atom
into the ground and various excited $S-$states of the Be$^+$ ion. The probability of ionization from Be$^+$ ion to Be$^{2+}$ ion
is calculated as $P_{ion}=1-\sum_{i=1}^{\infty} P_i$. }
\begin{center}
\scalebox{0.90}{%
\begin{tabular}{cccc}
\hline\hline
State of Be$^+$ & $\;\;$  Amplitude $\;\;$   & $\;\;$ Probability ($P_i$) $\;\;$ & $\;\;$  $P_i$  in $\%$ \\
\hline
 2$^2$S &  0.759 683 487 & 0.577 119  &  57.71   \\
 3$^2$S &  0.514 929 058 & 0.265 152  &  26.52   \\
 4$^2$S &  0.073 790 160 & 0.005 445  &   0.54   \\
 5$^2$S &  0.043 113 179 & 0.001 859  &   0.19   \\
 6$^2$S &  0.029 411 301 & 0.000 865  &   0.09   \\
 7$^2$S &  0.021 688 396 & 0.000 470  &   0.05   \\
Total   &                & 0.850 910  &  85.09   \\
$P_{ion}$&               & 0.149 090  &  14.91   \\
\hline\hline
\end{tabular}}
\end{center}
\end{table}


\begin{table}
\caption{Transition probabilities for the nuclear $\beta^-$ decay from the excited $3^2S$-, $4^2S$-, $5^2S$-,
$6^2S$-, and $7^2S$-states of the Li atom$^a$
into the ground and various excited states of the Be$^+$ ion$ $.}
\begin{center}
\scalebox{0.75}{%
\begin{tabular}{cccr}
\hline\hline
States Li $\rightarrow$ Be$^+$ & Amplitude  \qquad &  \qquad Probability &  $P_i$  in $\%$ \\
\hline
3$^2$S $\rightarrow$ 2$^2$S & 0.239 962 786  & 0.057 582  &  5.76 \\
3$^2$S $\rightarrow$ 3$^2$S & 0.466 529 800  & 0.217 650  & 21.76 \\
3$^2$S $\rightarrow$ 4$^2$S & 0.757 456 066  & 0.573 740  & 57.37 \\
3$^2$S $\rightarrow$ 5$^2$S & 0.055 586 071  & 0.003 090  &  0.31 \\
3$^2$S $\rightarrow$ 6$^2$S & 0.012 740 357  & 0.000 162  &  0.02 \\
3$^2$S $\rightarrow$ 7$^2$S & 0.013 723 711  & 0.000 188  &  0.02 \\
Total                       &                & 0.852 412  & 85.24 \\
$P_{ion}$                   &                & 0.147 588  & 14.76 \\
\hline
4$^2$S $\rightarrow$ 2$^2$S & 0.132 669 559  & 0.017 601  &  1.76 \\
4$^2$S $\rightarrow$ 3$^2$S & 0.236 587 524  & 0.055 974  &  5.60 \\
4$^2$S $\rightarrow$ 4$^2$S & 0.122 373 066  & 0.014 975  &  1.50 \\
4$^2$S $\rightarrow$ 5$^2$S & 0.828 124 464  & 0.685 790  & 68.58 \\
4$^2$S $\rightarrow$ 6$^2$S & 0.277 774 076  & 0.077 158  &  7.72 \\
4$^2$S $\rightarrow$ 7$^2$S & 0.007 347 388  & 0.000 054  &  0.01 \\
Total                       &                & 0.851 552  & 85.16 \\
$P_{ion}$                   &                & 0.148 447  & 14.84 \\
\hline
5$^2$S $\rightarrow$ 2$^2$S & 0.087 318 854  & 0.007 625  &  0.76 \\
5$^2$S $\rightarrow$ 3$^2$S & 0.148 984 137  & 0.022 196  &  2.22 \\
5$^2$S $\rightarrow$ 4$^2$S & 0.109 684 232  & 0.012 031  &  1.20 \\
5$^2$S $\rightarrow$ 5$^2$S & 0.175 864 858  & 0.030 928  &  3.09 \\
5$^2$S $\rightarrow$ 6$^2$S & 0.698 154 162  & 0.487 419  & 48.74 \\
5$^2$S $\rightarrow$ 7$^2$S & 0.503 067 106  & 0.253 076  & 25.31 \\
Total                       &                & 0.813 276  & 81.33 \\
$P_{ion}$                   &                & 0.186 724  & 18.67 \\
\hline
6$^2$S $\rightarrow$ 2$^2$S & 0.063 750 613  & 0.004 064  &  0.41 \\
6$^2$S $\rightarrow$ 3$^2$S & 0.104 007 178  & 0.010 817  &  1.08 \\
6$^2$S $\rightarrow$ 4$^2$S & 0.079 072 301  & 0.006 252  &  0.63 \\
6$^2$S $\rightarrow$ 5$^2$S & 0.071 415 619  & 0.005 100  &  0.51 \\
6$^2$S $\rightarrow$ 6$^2$S & 0.350 972 033  & 0.123 181  & 12.32 \\
6$^2$S $\rightarrow$ 7$^2$S & 0.430 715 551  & 0.185 516  & 18.55 \\
\hline\hline
\end{tabular}}
\end{center}
{\footnotesize
\footnotetext[1]{The calculated total energy of the first excited  S-state of lithium atom is -7.354 093 706 a.u.
($3^{2}S$-state), while the total energies of the second and higher excited states are
-7.318 517 759 a.u. ($4^{2}S$-state),
-7.303 458 818 a.u. ($5^{2}S$-state) and
-7.295 734 702 a.u. ($6^{2}S$-state), respecively.}
\footnotetext[2]{The total energies of the same $S-$states in the Be$^+$ can be found in Table I.}
}
\end{table}


\begin{table}
\caption{Transition probabilities between states of P-symmetry for the nuclear $\beta^-$-decay of the Li$^a$ atom to the Be$^+$ ion$^b$.}
\begin{center}
\scalebox{0.65}{%
\begin{tabular}{cccr}
\hline\hline
States Li $\rightarrow$ Be$^+$ & Amplitude \qquad & \qquad Probability  & $P_i$  in $\%$ \\
\hline
2$^2$P $\rightarrow$ 2$^2$P & 0.697 549 959 & 0.486 576 & 48.66 \\
2$^2$P $\rightarrow$ 3$^2$P & 0.603 885 572 & 0.364 678 & 36.47 \\
2$^2$P $\rightarrow$ 4$^2$P & 0.003 979 607 & 0.000 016 &  0.00 \\
2$^2$P $\rightarrow$ 5$^2$P & 0.020 232 690 & 0.000 409 &  0.04 \\
2$^2$P $\rightarrow$ 6$^2$P & 0.013 143 263 & 0.000 173 &  0.02 \\
2$^2$P $\rightarrow$ 7$^2$P & 0.013 285 358 & 0.000 176 &  0.02 \\
Total                       &               & 0.852 028 & 85.20 \\
$P_{ion}$                   &               & 0.147 972 & 14.80 \\
\hline
3$^2$P $\rightarrow$ 2$^2$P & 0.275 908 160 & 0.076 125 &  7.61 \\
3$^2$P $\rightarrow$ 3$^2$P & 0.319 479 925 & 0.102 067 & 10.21 \\
3$^2$P $\rightarrow$ 4$^2$P & 0.801 261 129 & 0.642 019 & 64.20 \\
3$^2$P $\rightarrow$ 5$^2$P & 0.166 010 974 & 0.027 560 &  2.76 \\
3$^2$P $\rightarrow$ 6$^2$P & 0.004 047 006 & 0.000 016 &  0.00 \\
3$^2$P $\rightarrow$ 7$^2$P & 0.004 025 567 & 0.000 016 &  0.00 \\
Total                       &               & 0.847 804 & 84.78 \\
$P_{ion}$                   &               & 0.152 196 & 15.22 \\
\hline
4$^2$P $\rightarrow$ 2$^2$P & 0.161 045 822 & 0.025 936 &  2.59 \\
4$^2$P $\rightarrow$ 3$^2$P & 0.195 960 248 & 0.038 400 &  3.84 \\
4$^2$P $\rightarrow$ 4$^2$P & 0.046 100 299 & 0.002 125 &  0.21 \\
4$^2$P $\rightarrow$ 5$^2$P & 0.724 469 360 & 0.524 856 & 52.49 \\
4$^2$P $\rightarrow$ 6$^2$P & 0.425 779 325 & 0.181 288 & 18.13 \\
4$^2$P $\rightarrow$ 7$^2$P & 0.425 325 535 & 0.180 902 & 18.09 \\
Total                       &               & 0.953 507 & 95.35 \\
$P_{ion}$                   &               & 0.046 493 &  4.65 \\
\hline
5$^2$P $\rightarrow$ 2$^2$P & 0.113 441 928 & 0.012 869 &  1.29 \\
5$^2$P $\rightarrow$ 3$^2$P & 0.135 765 197 & 0.018 432 &  1.84 \\
5$^2$P $\rightarrow$ 4$^2$P & 0.017 086 489 & 0.000 292 &  0.03 \\
5$^2$P $\rightarrow$ 5$^2$P & 0.328 135 053 & 0.107 673 & 10.77 \\
5$^2$P $\rightarrow$ 6$^2$P & 0.547 505 865 & 0.299 763 & 29.98 \\
5$^2$P $\rightarrow$ 7$^2$P & 0.550 438 371 & 0.302 982 & 30.30 \\
\hline
6$^2$P $\rightarrow$ 2$^2$P & 0.081 224 665 & 0.006 597 &  0.66 \\
6$^2$P $\rightarrow$ 3$^2$P & 0.099 158 755 & 0.009 832 &  0.98 \\
6$^2$P $\rightarrow$ 4$^2$P & 0.029 284 955 & 0.000 858 &  0.09 \\
6$^2$P $\rightarrow$ 5$^2$P & 0.177 208 892 & 0.031 403 &  3.14 \\
6$^2$P $\rightarrow$ 6$^2$P & 0.353 201 418 & 0.124 751 & 12.48 \\
6$^2$P $\rightarrow$ 7$^2$P & 0.348 353 690 & 0.121 350 & 12.13 \\
\hline\hline
\end{tabular}}
\end{center}
\footnotetext[1]{The total energies of the incident P-states of Li atom are: 
-7.410 149 067 a.u. ($2^2P-$state), 
-7.337 050 609 a.u.($3^2P-$state), 
-7.311 770 213 a.u.($4^2P-$state), 
-7.299 899 542 a.u. ($5^2P-$state) and 
-7.293 494 640 a.u. ($6^2P-$state), respectively.}
\footnotetext[2]{The total energies of the same (final) $P-$states of the Be$^+$ ion are: 
-14.179 326 074 a.u., -13.885 034 739 a.u., -13.783 519 845 a.u., -13.733 901 878 a.u., -13.711 935 225 a.u. and -13.711 378 665 a.u., respectively}
\end{table}


\begin{table}
\caption{Transition probabilities between states of D-symmetry for the nuclear $\beta^-$-decay of the Li atom$^a$ to the Be$^+$ ion$^b$.}
\begin{center}
\scalebox{0.65}{%
\begin{tabular}{cccr}
\hline\hline
States Li $\rightarrow$ Be$^+$ & Amplitude \qquad & \qquad Probability  & $P_i$  in $\%$ \\
\hline
3$^2$D $\rightarrow$ 3$^2$D & 0.613 879 768 & 0.376 848 & 37.68 \\
3$^2$D $\rightarrow$ 4$^2$D & 0.675 444 736 & 0.456 226 & 45.62 \\
3$^2$D $\rightarrow$ 5$^2$D & 0.124 573 183 & 0.015 518 &  1.55 \\
3$^2$D $\rightarrow$ 6$^2$D & 0.005 331 066 & 0.000 028 &  0.00 \\
3$^2$D $\rightarrow$ 7$^2$D & 0.008 233 705 & 0.000 068 &  0.01 \\
3$^2$D $\rightarrow$ 8$^2$D & 0.004 516 156 & 0.000 020 &  0.00 \\
Total                       &               & 0.848 709 & 84.87 \\
$P_{ion}$                   &               & 0.151 291 & 15.13 \\
\hline
4$^2$D $\rightarrow$ 3$^2$D & 0.297 395 858 & 0.088 444 &  8.84 \\
4$^2$D $\rightarrow$ 4$^2$D & 0.092 613 446 & 0.008 577 &  0.86 \\
4$^2$D $\rightarrow$ 5$^2$D & 0.645 316 813 & 0.416 434 & 41.64 \\
4$^2$D $\rightarrow$ 6$^2$D & 0.314 342 441 & 0.098 811 &  9.88 \\
4$^2$D $\rightarrow$ 7$^2$D & 0.002 237 872 & 0.000 005 &  0.00 \\
4$^2$D $\rightarrow$ 8$^2$D & 0.000 807 365 & 0.000 001 &  0.00 \\
Total                       &               & 0.612 272 & 61.23 \\
$P_{ion}$                   &               & 0.387 728 & 38.77 \\
\hline
5$^2$D $\rightarrow$ 3$^2$D & 0.221 060 550 & 0.048 868 &  4.89 \\
5$^2$D $\rightarrow$ 4$^2$D & 0.131 551 368 & 0.017 306 &  1.73 \\
5$^2$D $\rightarrow$ 5$^2$D & 0.303 521 357 & 0.092 125 &  9.21 \\
5$^2$D $\rightarrow$ 6$^2$D & 0.662 141 399 & 0.438 431 & 43.84 \\
5$^2$D $\rightarrow$ 7$^2$D & 0.408 116 762 & 0.166 559 & 16.66 \\
5$^2$D $\rightarrow$ 8$^2$D & 0.051 512 344 & 0.002 653 &  0.26 \\
Total                       &               & 0.765 943 & 76.59 \\
$P_{ion}$                   &               & 0.234 057 & 23.41 \\
\hline
6$^2$D $\rightarrow$ 3$^2$D & 0.219 012 937 & 0.047 966 &  4.80 \\
6$^2$D $\rightarrow$ 4$^2$D & 0.109 576 684 & 0.012 007 &  1.20 \\
6$^2$D $\rightarrow$ 5$^2$D & 0.275 765 182 & 0.076 046 &  7.60 \\
6$^2$D $\rightarrow$ 6$^2$D & 0.172 623 479 & 0.029 799 &  2.98 \\
6$^2$D $\rightarrow$ 7$^2$D & 0.238 431 045 & 0.056 849 &  5.68 \\
6$^2$D $\rightarrow$ 8$^2$D & 0.469 886 268 & 0.220 793 & 22.08 \\
\hline
7$^2$D $\rightarrow$ 3$^2$D & 0.250 505 561 & 0.062 753 &  6.28 \\
7$^2$D $\rightarrow$ 4$^2$D & 0.224 414 920 & 0.050 362 &  5.04 \\
7$^2$D $\rightarrow$ 5$^2$D & 0.113 445 739 & 0.012 870 &  1.29 \\
7$^2$D $\rightarrow$ 6$^2$D & 0.227 483 595 & 0.051 749 &  5.17 \\
7$^2$D $\rightarrow$ 7$^2$D & 0.476 006 811 & 0.226 582 & 22.66 \\
7$^2$D $\rightarrow$ 8$^2$D & 0.429 479 022 & 0.184 452 & 18.45 \\
\hline\hline
\end{tabular}}
\end{center}
{\footnotesize
\footnotetext[1]{The total energies of the incident D-states of Li atoms are: -7.335 511 694 a.u ($3^2D$-state), 
-7.311 211 047 a.u. ($4^2D$-state), -7.298 835 884 a.u. ($5^2D$-state), -7.288 077 393 a.u. ($6^2D$-state) and 
-7.268 731 551 a.u. ($7^2D$-state)}
\footnotetext[2]{The total energies of the same (final) states of the Be$^+$ ion are: 
-13.878 004 697 a.u., -13.778 986 828 a.u., -13.733 832 498 a.u., -13.705 903 173 a.u., -13.677 409 085 a.u. and 
-13.660 271 947 a.u.}}
\end{table}


\begin{thebibliography}{99}
\bibitem{Our1}  A.M. Frolov and M.B. Ruiz, ''Atomic excitations during the nuclear $\beta^-$ decay 
in light atoms'', Phys. Rev. A \textbf{82}, 042511 (2010).

\bibitem{FroTal}  A.M. Frolov and J.D. Talman, ''Final-state probabilities for $\beta^-$-decaying 
light atoms'', Phys. Rev. A \textbf{72}, 022511 (2005).

\bibitem{Mig}  A.B. Migdal, ''Ionization of atoms accompanying $\alpha$- and $\beta$-decay'', 
J. Phys. Acad. Sci. USSR 4(1-6), 449-453 (1941).

\bibitem{MigK}  A.B. Migdal and V. Krainov, \textit{Approximation Methods in
Quantum Mechanics.}, (W.A. Benjamin, New York (1969)).

\bibitem{LLQ}  L.D. Landau and E.M. Lifshitz, \textit{Quantum Mechanics:
non-relativistic theory.}, (3rd. ed. Pergamon Press, New York (1976)), Chpt.
VI.

\bibitem{Frolov-Li}  A.M. Frolov, ''Compact and accurate variational wave functions of three-electron 
atomic systems constructed from semi-exponential radial basis functions'', Eur. Phys. J. D \textbf{61}, 571-577 (2011).

\bibitem{Larsson}  S. Larsson, ''Calculations on the $^2$S ground state of the lithium atom using wave functions of Hylleraas type''
Phys. Rev. \textbf{169}, 49-54 (1968).

\bibitem{Dir}  P.A.M. Dirac, \textit{The Principles of Quantum Mechanics.},
(4th ed.,Oxford at the Clarendon Press, Oxford (1958)).

\bibitem{FrWa2010}  A.M. Frolov and D.M. Wardlaw, ''On bound state computations in three- and four-electron atomic systems'', 
JETP \textbf{138}, 5-15 (2010).

\bibitem{Fro2012}  A.M. Frolov, ''On the nuclear $(n;t)-$reaction in
the three-electron Li atom'', ArXiv: 1207.3471 [phys.,at.-phys.] (2012).

\bibitem{GR}  I.S. Gradstein and I.M. Ryzhik, \textit{Tables of Integrals,
Series and Products}, (6th revised ed., Academic Press, New York, (2000)).

\bibitem{SimsJCP}  J.S. Sims and S.A. Hagstrom, ''One-center $r_{ij}$ integrals over Slater orbitals'', J. Chem. Phys. \textbf{55},
4699-4710 (1971).

\bibitem{Sims-Be}  J.S. Sims and S.A. Hagstrom, ''Combined Configuration-interaction-Hylleraas-type wave-function study of the 
ground state of the beryllium atom'', Phys. Rev. A \textbf{4}, 908-916 (1971).

\bibitem{Our3}  M.B. Ruiz, J.T. Margraf, and A.M. Frolov, ''Hylleraas-configuration-interaction analysis of the low-lying 
states in the three-electron Li atom and Be$^+$ ion, submitted (2013).

\bibitem{Pachucki}  M. Puchalski, D. Kedziera, and K. Pachucki, ''Ground state of Li and Be$^+$ using explicitly correlated functions'', 
Phys. Rev. A \textbf{80}, 032521 (2009).

\bibitem{PP} M. Puchalski and K. Pachucki, ''Relativistic, QED, and finite nuclear mass corrections for low-lying states of Li and Be$^+$'',  
Phys. Rev. A {\bf 78} 052511 (2008).

\bibitem{Drake-Li}  L.M. Wang, Z.-C. Yan, H.X. Qiao, and G.W.F. Drake, ''Variational energies and the Fermi contact term for the low-lying states 
of lithium: Basis-set completeness'', Phys. Rev. A \textbf{85}, 052513 (2012).

\bibitem{Sims-Li}  J.S. Sims and S.A. Hagstrom, Hylleraas-configuration-interaction study of the 2$^2$S ground state of neutral lithium 
and the first five excited $^2$S states'', Phys. Rev. A \textbf{80}, 052507 (2009).

\bibitem{King-Li}  F.W. King, ''High-precision calculations for the ground and excited states of the lithium atom'',  
Adv. At. Mol. Opt. Phys. \textbf{40}, 57-112 (1999). 

\bibitem{Thakkar}  A.J. Thakkar, T. Koga, T. Tanabe, and H. Teruya, Chem.
Phys. Lett. \textbf{366}, 95 (2002).

\bibitem{Ruiz2e}  M.B. Ruiz, '' Evaluation of Hylleraas-CI atomic integrals. III. Two-electron kinetic energy integrals'', 
J. Math. Chem. \textbf{49}, 2457-2485 (2011).

\bibitem{Ruiz3e}  M.B. Ruiz, '' Evaluation of Hylleraas-CI atomic integrals by integration over the coordinates of one electron. 
I. Three-electron integrals'', J. Math. Chem. \textbf{46}, 24-64 (2009). 

\bibitem{Sims-triang}  J.S. Sims and S.A. Hagstrom, ''Mathematical and computational science issues in high precision Hylleraas-configuration 
interaction variational calculations: I. Three-electron integrals'', J. Phys. B: At. Mol. Opt. Phys. \textbf{37}, 1519-1540 (2004).

\bibitem{2beta} A. Giuliani and A. Poves, "Neutrinoless Double-Beta Decay". Advances in High Energy Physics (2012): doi:10.1155/2012/857016.

\bibitem{Adam-Be+} M. Stanke, J. Komasa, D. Kedziera, S. Bubin, and L. Adamowicz, ''Three lowest S states of $^9$Be$^+$ calculated with including 
nuclear motion and relativistic and QED corrections'', 
Phys. Rev. A \textbf{77}, 062509 (2008).

\bibitem{King-Be+} F.W. King, ''High-precision calculations of the hyperfine constants for the low-lying excited $^2$S states of Be$^+$, 
J. Phys. Chem. A \textbf{113}, 4110-4116 (2009).


\end{thebibliography}
\end{document}